%% file: main.tex
\documentclass{pasa}%
\pdfoutput=1
\usepackage{subcaption}
\usepackage{graphicx}
\usepackage{physics}
\usepackage{gensymb}
\usepackage{wasysym}
\usepackage{blindtext}

\title[A Low Earth Orbit Astronomical Interferometer]{A Linear Formation Flying Astronomical Interferometer in Low Earth Orbit}

\author[Hansen et al.]{Jonah T. Hansen$^1$, Michael J. Ireland$^1$
\affil{$^1$Research School of Astrononmy \& Astrophysics, Australian National University, ACT 2611, Australia}%
}%

\jid{PASA}
\jyear{\the\year}

\usepackage{aas_macros}
\usepackage{hyperref}
\hypersetup{colorlinks,citecolor=blue,linkcolor=blue,urlcolor=blue}

\newcommand*\chem[1]{\ensuremath{\mathrm{#1}}}
\newcommand{\red}{\color{black}}

\hypersetup{draft}

\begin{document}

\begin{frontmatter}
\maketitle

\begin{abstract}
Space interferometry is the inevitable endpoint of high angular resolution astrophysics, and a key technology that can be leveraged to analyse exoplanet formation and atmospheres with exceptional detail. However, the anticipated cost of large missions such as Darwin and TPF-I, and inadequate technology readiness levels have resulted in limited developments since the late 2000s. Here, we present a feasibility study into a small scale formation flying interferometric array in Low Earth Orbit, that will aim to prove the technical concepts involved with space interferometry while still making unique astrophysical measurements. We will detail the proposed system architecture and metrology system, as well as present orbital simulations that show that the array should be stable enough to perform interferometry with < 50m/s/year delta-v and one thruster per spacecraft. We also conduct observability simulations to identify which parts of the sky are visible for a given orbital configuration. We conclude with optimism that this design is achievable, but a more detailed control simulation factoring in a demonstrated metrology system is the next step to demonstrate full mission feasibility.
\end{abstract}

\end{frontmatter}

\input{1Intro}
\input{2SystemArchitecture}
\input{3OrbitalDynamics}
\input{4Conclusion}

\section*{Acknowledgements}

{\red The authors would like to thank Tiphaine Lagadec and Harry-Dean Kenchington Goldsmith for comments on the manuscript.}

\bibliographystyle{pasa-mnras}
\bibliography{references,references_mike}

\end{document}

%% file: 1Intro.tex
\section{Introduction}
\label{sec:intro}
\subsection{The Case for Space Interferometry}

Since the first confirmed exoplanet detection of 51 Peg b by \cite{Mayor1995}, there has been a desire to directly image and study exoplanets without relying on indirect methods such as transits and radial velocity measurements. While these methods have produced a large number of exoplanet detections since 51 Peg b, they are limited by selection biases and do not allow many of the properties of the planet to be studied. In particular, one of the major goals of exoplanet research is identifying potentially habitable worlds that may harbour life. To determine this, one would need to study the atmosphere of the planet. Transmission spectroscopy {\red is especially promising for characterisation for hydrogen rich atmospheres, but is challenging for terrestrial atmospheres \citep[e.g.][]{DiamondLowe19}. The rich data sets expected from missions such as ARIEL \citep{Tinetti18} will also focus on likely tidally locked locked planets around small stars. A knowledge of atmospheres in system types that bracket the earth-sun system require a broader suite of techniques.}

This is where direct imaging becomes important - by directly imaging an exoplanet, it is possible to obtain information about the atmosphere and surface, and therefore determine if a ``biosignature'' is present. According to \cite{Leger2011}, a biosignature is ``an observable feature of a planet, such as its atmospheric composition, that our present models cannot reproduce when including the abiotic physical and chemical processes we know about''. Biosignatures were first studied by \cite{Sagan1993}, who used the Galileo probe to look back at Earth and determine if a biosignature could be found \textit{a priori}. This study was successful, finding that a multitude of molecules such as \chem{O_2} were significantly out of thermodynamical equilibrium and were inexplicable through abiotic processes. Extending this analysis to exoplanets is a tantalising research goal.

However, many obstacles stand in the way of direct imaging of an exoplanet. The most prominent of these concerns separating the emission of the planet from that of its host star. The contrast between the star and a 300\,K terrestrial planet (using the Sun/Earth ratio as an analogue) is an astounding $10^{10}$ in visible light, but viewing in the mid-infrared (10\,$\mu$m) reduces the contrast to $10^7$ \citep{Angel1997}. There are other reasons for choosing the mid-infrared as well - {\red in particular the presence of strong molecular biosignature tracers. 
Some specific lines of interest include \chem{CH_4} at 7.7\,$\mu$m, \chem{O_3} at 9.7\,$\mu$m, \chem{N_2O} at 7.8\,$\mu$m and \chem{CO_2} at 15\,$\mu$m. 
\chem{O_3} is of particular note as it can be used as a strong tracer for \chem{O_2}. A fuller treatment on molecular biosignatures can be found \cite[e.g.][]{Schwieterman18,Quanz2019AtmosphericDiversity}}.

While the mid-infrared contains many advantages over the visible, there are still two main complications. The first issue is the substantial background from ground based telescopes, requiring space-based telescopes located far from the earth to achieve zodiacal-limited observations. The second issue concerns the baselines required to achieve high angular resolution: for a planet 1\,AU out from a host star at 10\,pc, {\red a minimum angular resolution of 0.1\,arcsec would be required to distinguish the planet from the star. For a coronagraphic inner working angle of 2 $\lambda$/D and a 15 micron wavelength, detecting such an exoplanet would require a 60m telescope}. With interferometry, one can achieve high spatial resolution with small apertures, {\red circumnavigating the issues of a monolithic telescope with such a resolution requirement and especially for space-based telescopes \citep{Angel1997}}. A mid-infrared interferometer would be complementary to an optical/near-infrared coronagraph space telescope, and would detect more planets around cool stars in particular \citep{Quanz2018ExoplanetInterferometer}. This is significant as the costs for a large scale space-based mid-infrared interferometric mission (such as \textit{Darwin}) would be more economical ($\sim$\$1.2B \citep{Cockell2009DarwinanPlanets}) than the equivalent space-based optical telescope (e.g. LUVIOR at $>$ \$10B \citep{NationalAeronauticsandSpaceAdministration2019LUVIORReport}).

Although this is arguably the most compelling case for space interferometry, the joint requirements for sensitivity and angular resolution mean that there are a great number of additional key science cases. These include direct detection of forming exoplanets \citep{Monnier18}, imaging cool gas and dust in the far infrared \citep{Leisawitz07}, and measuring black hole masses at cosmological distances \citep{Gravity18}

\subsection{Developments in Space-based Interferometry}

The first design of a space-based interferometer for the purpose of exoplanet research was proposed by \cite{Bracewell1978}. His design, known as a nulling interferometer, was the basis for most of the interferometers designed in future years. A nulling interferometer is designed to enhance the planet's light over the power of the star by placing an interference null over the star. This is achieved by having two incident beams of light with the same optical path length and introducing a phase shift of $\pi$ on one of them \citep{Fridlund2004}. The light then destructively interferes at the central minimum and constructively interferes at an angle $\theta = \lambda/2\Delta b$ away from the central minimum. If the baseline of the interferometer is adjusted so that the planet is an angle $\theta$ away from the star, then the planet's emission is maximised while most of the star's light is removed. The contrast is then limited to light leaks from the the central minimum due to optical and mechanical imperfections \citep{Fridlund2004}. This raw contrast only needs to be better than the zodiacal background: calibration in the presence of background fluctuations and fringe tracking errors requires at least 3 telescopes and a so-called ``kernel-nulling'' configuration \citep{Martinache2018Kernel-nullingPlanets}.

Space interferometry was studied extensively in the 1980s and 1990s, particularly by the European Space Agency (ESA) who identified it as a key goal in the Horizon 2000 plan \citep{Fridlund2002}. As a result of this, many proposals and concepts for missions were developed over several conferences and workshops. Among these were rigid structures such as COSMIC \citep{Traub1985} and FLUTE \citep{Labeyrie1980}, and free flying interferometers such as OASIS \citep{Noordam1985}, TRIO \citep{Labeyrie1985} and SAMSI \citep{Stachnik1985}. The last two are of particular note. The TRIO proposal eventually led to the development of the \textit{Darwin} mission \citep{Leger1996,LabeyrieA.SchneiderJ.BoccalettiA.RiaudP.MoutouC.AbeL.Rabou2000}. The other proposal, SAMSI, shares some details in common with the interferometric array that we will investigate in this paper. 

From these initial proposals, two flagship missions were devised: NASA's Terrestrial Planet Finder Interferometer (TPF-I) mission \citep{Beichman1999} and ESA's \textit{Darwin} mission \citep{Leger1996}. Both of these missions were free flying arrays of satellites orbiting around the Sun-Earth L2 point. 

TPF-I aimed to study terrestrial exoplanets using four 3.5\,m telescopes and one central spacecraft holding the beam combiner, observing at wavelengths between 7-20\,$\mu$m \citep{Beichman1999}. The particular arrangement was originally the double Bracewell proposed by \cite{Angel1997}, but eventually became an ``Emma X-array'', an arrangement consisting of the four telescopes flying in rectangular formation and the beam combiner located about 1200\,m above the array \citep{Defrere2018}. Two pathfinder missions were initially identified to demonstrate the technology required for TPF-I: the Space Interferometry Mission SIM \citep{Shao1998} and the Starlight mission \citep{Blackwood2003}. However neither of the pathfinders, and eventually TPF-I itself, were funded due to budget cuts in 2007. 

Similarly, \textit{Darwin} was originally designed to have six 1.5\,m satellites in a hexagonal configuration, operating between $5-10\,\mu$m, with a beam combiner in the middle \citep{Fridlund2004}. This configuration was designed to allow the use of multiple baselines for better nulling \citep{Angel1997} and to allow for two dimensional imaging of an exoplanet system. The baselines were designed to be tuned to an individual star so that the habitable zone was always at the angle of maximal constructive interference, a baseline typically around 1\,km. In order to distinguish the planet signal from the exozodiacal light, the whole system was designed to be rotated and switched asymmetrically between different orientations, so that the signal was modulated. Eventually, \textit{Darwin} also adopted a four telescope/one beam combiner ``Emma X-array'' configuration near the end of the design study \citep{Cockell2009DarwinanPlanets}. Again, due to budget cuts, \textit{Darwin} was not funded and ceased being investigated in 2007. ESA is currently planning for the next 30 years as part of the Voyage 2050 process, and a revised \textit{Darwin}-like mission LIFE (Large Interferometer For Exoplanets) (\cite{Quanz2018ExoplanetInterferometer}; \cite{Quanz2019AtmosphericDiversity}) is part of the discussion.

Since the demise of the \textit{Darwin} and TPF-I missions, there has been limited progress on space interferometry. Here, we propose a pathway towards the eventual goal of an exoplanet imaging interferometer; specifically, a linear formation flying interferometer in Low Earth Orbit (LEO), designed as a technological pathfinder towards this end goal. This array will consist of three CubeSat satellites: two 3U CubeSats which will act as telescopes, and one 6U beam combiner located between the two. 
The interferometer is planned to operate in the visible, due to the difficulty in cooling a telescope in LEO, and the additional power and mass required to cool an infrared sensor. 

In this paper, we will propose the system architecture of the satellite system, including the associated metrology architecture and requirments. We will then outline the orbital configuration for such a mission, as well as a treatment of the relative perturbations associated with LEO. Finally, we will discuss observability of astrophysical targets and potential science foci for this class of mission.

%% file: 2SystemArchitecture.tex
\section{System Architecture}
\subsection{Array Design}
\label{refArrayDesign}

\begin{figure*}[t!]
    \centering
    \includegraphics[width=0.6\linewidth]{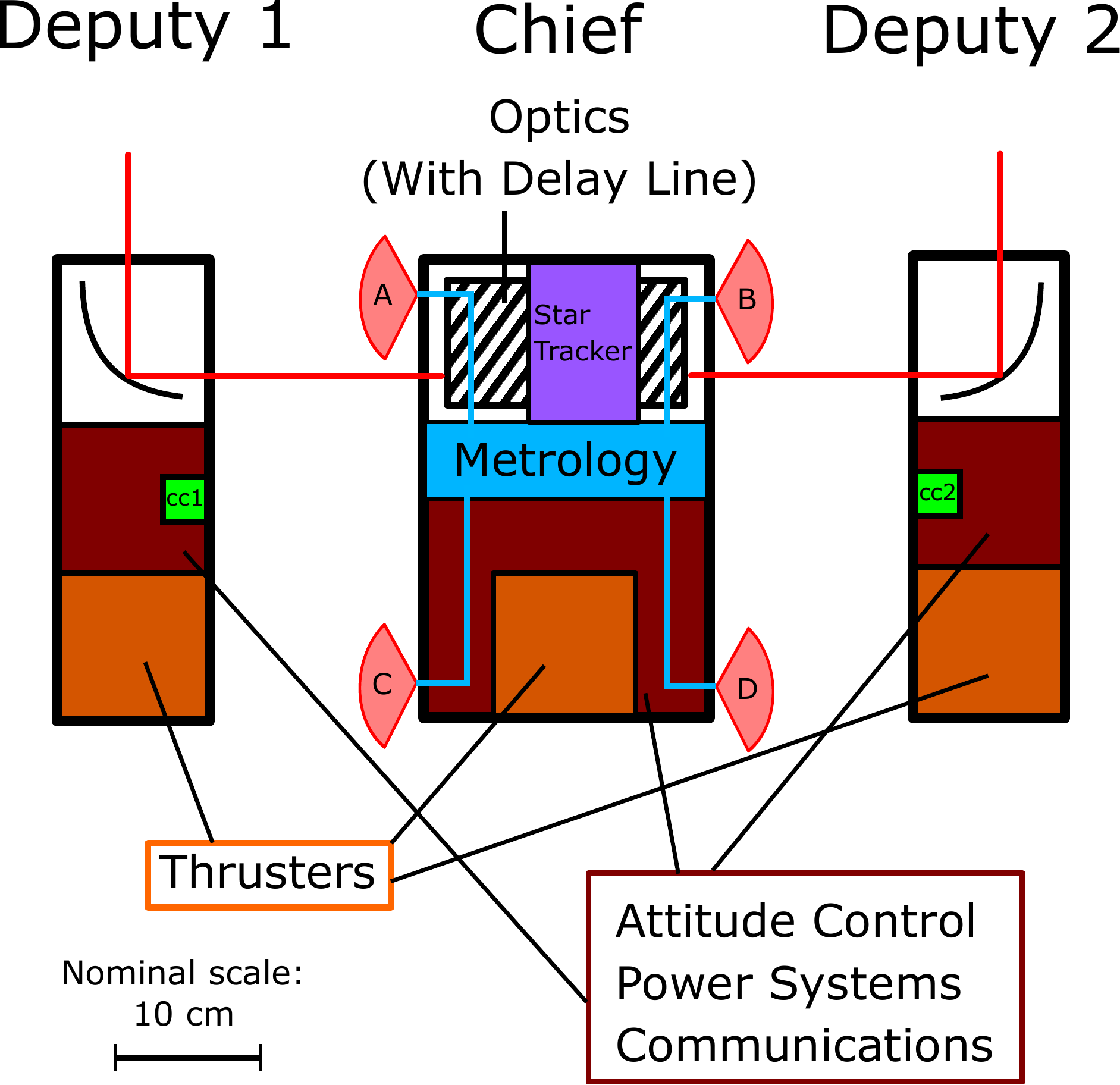}
    \caption{Schematic of the satellite array. The metrology system measures distances from optical heads A and C to corner cube 1 (cc1), and from optical heads B and D to corner cube 2 (cc2).}
    \label{asp_diagram}
\end{figure*}

Our proposed design is to have a linear array of three nano-satellites: two telescopes (hereby called the deputies) and one beam-combiner (called the chief). The telescopes would ideally be 3U CubeSats to minimise cost, containing 2-3 mirrors of the optical system, solar power system, attitude control and a single thruster. The beam-combiner would be larger: a 6U CubeSat with the required interferometric optics (including a small delay line), one thruster, power system, metrology system, attitude control and communication system. A diagram of this initial design, with required space for each subsystem, is shown in Figure~\ref{asp_diagram}. This is admittedly an ambitious design in terms of space requirements. We also anticipate that this mission will have an altitude of about 500\,km and a lifetime of three years.

To comfortably keep track of interference fringes, we need to keep the optical path between the two arms of the interferometer within the coherence length of the beam-combiner. The optical path difference is given by the equation
\begin{align}
    {\red \Lambda} = |\Delta\vb*{r}_1| - \Delta\vb*{r}_1\vdot\vu*{s} - |\Delta\vb*{r}_2| + \Delta\vb*{r}_2\vdot\vu*{s} \approx 0,
\label{eqnOPD}
\end{align}
where $\Delta\vb*{r}$  is the separation of one of the deputies from the chief, and $\vu*{s}$ is the unit vector in the direction of the star. To resolve velocities of 100\,km/s per spectral channel with a central {\red science} wavelength of ${\red\lambda_s}\sim$650\,nm, we have to distinguish wavelengths separated by
\begin{equation*}
    {\red \Delta \lambda_s = \lambda_s}\frac{v}{c} = 0.2\text{\,nm}.
\end{equation*}
 This results in a science detector resolving power of $R_s=3000$, and a coherence length of about 2\,mm, which is the required positional knowledge. However, an optical delay line inside the beam combiner should be able to alleviate the requirements on the satellites themselves, as was proposed for the \textit{Darwin} mission \citep{Karlsson2003TheDARWIN}. If the 6U form factor is adopted, then a delay line of several cm can be easily accommodated, resulting in position control at the few cm level.  

\subsection{Metrology}

Metrology for a formation flying interferometer is more challenging than ground-based interferometry, because there is no stable reference frame. There are two key requirements on the knowledge of the optical path difference {\red$\Lambda$} (Equation~\ref{eqnOPD}) for each baseline - that the path length is within the coherence length of the science beam combiner, and that the pathlength does not blur the fringes within an exposure time. These requirements can be written as:
\begin{align}
{\red \Lambda} &< 0.5 \lambda_s R_s  \\
{\red \frac{d\Lambda}{dt}} &< 0.2 \frac{\lambda_s}{\red{\Delta T}},
\end{align}
for a science combiner resolving power $R_s$, science combiner wavelength $\lambda_s$ and science combiner exposure time {\red $\Delta T$}. For a visible light combiner with 10\,s exposure times and $R_s=100$, these requirements translate to ${\red\Lambda}<25\,\mu$m and ${\red d\Lambda/dt}<10\,$nm/s. Note that these are not requirements for stability, but a requirement for a measurement of {\red$\Lambda$}. We will assume here that once {\red$\Lambda$} is known, an appropriate delay line can make the corrections. 

We will not consider faint star, externally phase-referenced metrology here, and will consider on-axis observations only. In this case, the chief satellite can contain a narrow field-of-view tip/tilt sensor with a circular aperture of diameter $D_\text{tt}$, and a shot-noise limited centroid precision {\red integrated over the full science exposure time} of \citep{Ireland20}:
\begin{align}
{\red\sigma_\text{tt} = 0.33" \times D_\text{tt}^{-2} \eta_\text{tt}^{-1/2} \lambda_\text{tt}^{3/2} \Delta \lambda_\text{tt}^{-1/2} 10^{0.2 m_{\lambda,\text{tt}}} \Delta T^{-1/2}.}
\end{align}
This is shown as the "Star Tracker" in Figure~\ref{asp_diagram}. For example, for ${\red D_\text{tt}}=0.1\,$m, ${\red\lambda_\text{tt}}=0.5\,\mu$m, ${\red\Delta \lambda_\text{tt}}=0.3\,\mu$m, efficiency ${\red\eta_\text{tt}}=0.5$ and AB magnitude ${\red m_{\lambda,\text{tt}}}=10$, the centroid precision is $10^{-3}$ arcsec in 10\,s. In this example, the sensor pixel scale should be smaller than 0.5\,arcsec per pixel and the field of view at least of order 20\,arcsec in order to guarantee acquisition within the accuracy of a standard star tracker. The metrology system can only measure positions with respect to the reference frame created by this tip/tilt sensor. The uncertainty in optical path length ${\red\Lambda}$ and its derivative due to this chief spacecraft tilt error is simply given by:
\begin{align}
\sigma_\text{{\red$\Lambda$}, tt} &= B \sigma_\text{tt} \nonumber \\ 
\sigma_\text{{\red d$\Lambda$/dt}, tt} &= 3.46 B \sigma_\text{tt} \Delta T^{-1} .
\end{align}
{\red where $B$ is the total baseline between the two deputies. We have assumed that the tip/tilt sensor makes many small exposures within the time of the science exposure, enabling the slope in $\Lambda$ to be calculated.} 

As an example, for the parameters given above and a {\red600\,m baseline, $\sigma_\text{d$\Lambda$/dt, tt}$ is 1\,$\mu$m/s, or 35\,nm/s if increasing $\Delta T$ to the practical maximum for a fringe acquisition stage of 5\,minutes}. The reference frame of the chief satellite does not need to be {\em stable} over this long period, but only stable enough for the distance measuring metrology to make accurate measurements. The distance measuring metrology needs to measure ${\red\Lambda}$ synchronised with the tip/tilt reference frame measurement. Using the metrology system as shown in Figure~\ref{asp_diagram}, we can estimate ${\red\Lambda}$ {\red to first order in differences like $(d_A - d_C)/(d_A + d_C)$ and $h/d_A$} as:
\begin{align}
{\red\Lambda} &= (d_B-d_D) \frac{d_B+d_D}{2h} + (d_A-d_C) \frac{d_A+d_C}{2h} + \nonumber \\ 
&\frac{(d_B-d_A) + (d_D-d_C) + (d_B-d_C) + (d_D-d_A)}{4}  \nonumber \\ 
& + \sin(\theta) \frac{d_A + d_B +d_C + d_D}{2},
\end{align}
where {\red $d$ is the distance between a sensor and the corner cube, and $h$ is the distance along the star vector position between sensors $A$ and $C$ (or equivalently between $B$ and $D$ as these should be identical)}. The absolute distances such as $d_A$ need to be measured and act as scaling factors for error signals like $(d_B-d_D)$ and $\sin(\theta)$, but the highest accuracy by far is needed in distances  $d_B-d_D$ and $d_A-d_C$. For example, with a {\red600\,m} baseline and $h=0.3$\,m, measuring ${\red\Lambda}$ to within 25\,$\mu$m requires measuring $d_B-d_D$ and $d_A-d_C$ to within {\red12.5\,nm}. The goal velocity precision of 10\,nm/s requires measuring the rate of change of $d_B-d_D$ and $d_A-d_C$ to {\red5\,pm/s}. This is challenging, and requires an interferometric technique. One example of such a system is the MSTAR developed at JPL \citep{Lay03}, although there are a range of multiple wavelength metrology architectures \citep[e.g.][]{Kok13} that could measure these distance differences interferometrically. We suggest that the difference sums could be measured by time of flight methods, as the absolute distances do not need to be known to high accuracy.

%% file: 3OrbitalDynamics.tex
\section{Orbital Dynamics}

\subsection{Orbital configuration}

For a space interferometer to function, we require multiple telescopes all pointing towards the same source with a minimal optical path difference (Equation~\ref{eqnOPD}), which can be achieved by ensuring that the array is perpendicular to the direction of incoming light ($\vu*{s}$). This highlights one of the main issues with a fixed interferometric array in LEO; a fixed array would require a large amount of thrust to maintain this, in the presence of tidal accelerations. These accelerations are of order $\mu r^{-3} |\Delta\vb*{r}| = \omega^2 |\Delta\vb*{ r}|$, with $\omega$ the angular velocity of the orbit, $\mu$ the Earth's standard gravitational parameter, $r$ the orbital radius and $|\Delta\vb*{r}|$ the satellite separation. For a 500\,km orbit and 300\,m spacecraft separation, this  is 0.4\,mm\,s$^{-2}$, which is beyond typical accelerations from plasma thrusters.

The issue is solved by using a free flying array of two telescopes and a beam combiner. Each satellite can be put into orbits of slightly different inclinations, such that all three spacecraft form a baseline that is constantly perpendicular to the star vector. In this manner, the perturbations only need to be corrected along the $\vu*{s}$ direction and the baseline direction; less thrust is required.  The nature of having satellites in orbits with different orbital planes means that the baseline would rotate along the orbit, providing better coverage of the UV plane, and the free flying nature of the array allows for multiple baseline lengths through reconfiguration. 

In order to control Equation~\ref{eqnOPD} without large angular mirror actuation, we need an orbital configuration of the array such that the two optical paths are perpendicular to the star vector at all times. That is, $\Delta\vb*{r}(t)\vdot\vu*{s} = 0$ for all $t$. This configuration, assuming no perturbations, is as follows. Note that we shall refer to the orbit of the beam combiner (or chief) satellite as the reference orbit.

\begin{figure*}
  \centering
  \begin{subfigure}{0.49\textwidth}
    \centering
    \includegraphics[width=\linewidth]{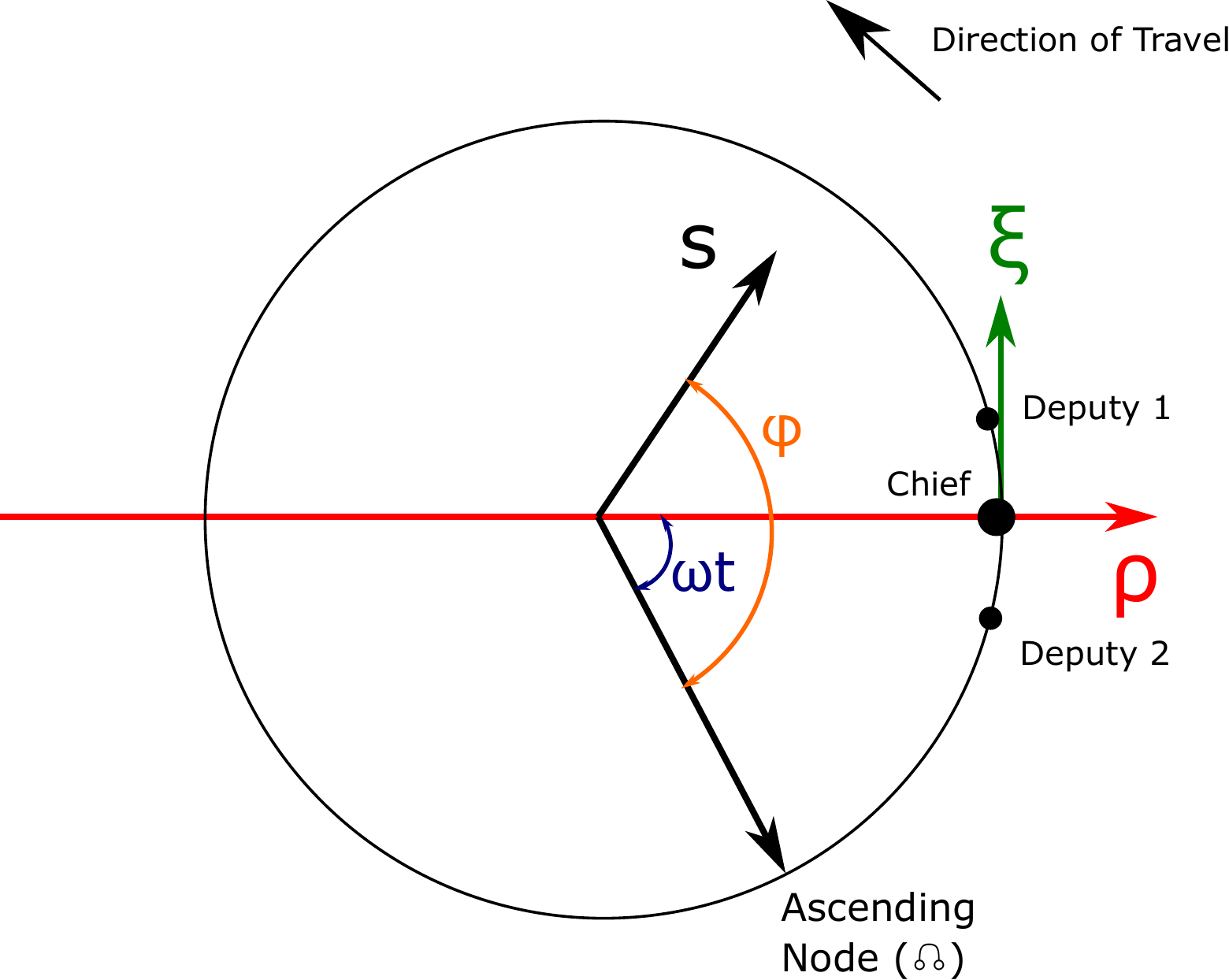}
    \caption{$\vu*{\rho}$-$\vu*{\xi}$ plane}
    \label{comb_X}
  \end{subfigure}
  \hfill
  \begin{subfigure}{0.49\textwidth}
    \centering
    \includegraphics[width=0.7\linewidth]{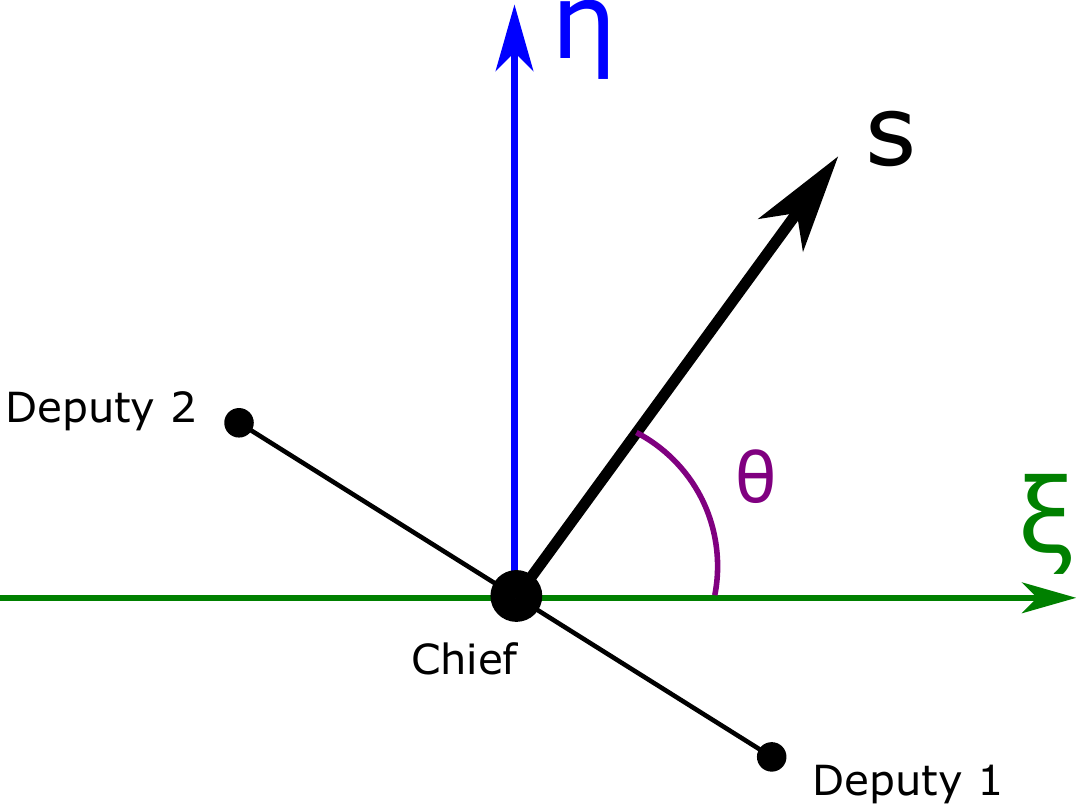}
    \caption{$\vu*{\xi}$-$\vu*{\eta}$ plane}
    \label{comb_y}
  \end{subfigure}
  \caption{Schematic of the LVLH frame, including the star vector angles $\theta$ and $\phi$.}
  \label{LVLH}
\end{figure*}

The reference orbit can be configured to be in any orbital plane; that is, with any inclination ($i$) and longitude of the ascending node ($\Omega$). From this reference orbit, there is a single configuration of the two deputy satellites that allows for the perpendicularity requirement of the array throughout the orbit. Consider the {\red local vertical, local horizontal (LVLH)} frame of reference; a rotating curvilinear frame with origin at the chief satellite, following its motion. The frame's unit vectors are defined as $\vu*{\rho}$ being along the radial position of the chief satellite, $\vu*{\xi}$ pointing in the along-track direction (velocity) of the chief satellite and $\vu*{\eta}$ pointing in the cross-track (angular momentum) of the chief satellite. Hence the orbital plane is equivalent to the $\vu*{\rho}-\vu*{\xi}$ plane.  A schematic of this reference frame is in Figure~\ref{LVLH}

Then, consider a star vector in this frame with spherical coordinates ($\theta,\phi$), where $\theta$ is the polar angle from the $\vu*{\eta}$ axis and $\phi$ is the azimuthal angle from the ascending node of the orbital plane (\ascnode) with respect to the equator 
\footnote{These two angles can be formed from the 
inclination ($i$) and longitude of the ascending node ($\Omega$) of the reference orbit, and right ascension ($\alpha$) and declination ($\delta$) of the target star as follows:
\begin{align*}\cos(\theta) &=\sin(\delta)\cos(i) - \cos(\delta)\sin(i) \sin(\alpha - \Omega)\\
\sin(\phi) &= \frac{\cos(\delta)\cos(i)\sin(\alpha-\Omega) + \sin(\delta)\sin(i)}{\sin(\theta)}\\
\cos(\phi) &=\frac{\cos(\delta) \cos(\alpha - \Omega)}{\sin(\theta)} \end{align*}}. 
Finally, suppose that a satellite in this orbit has a phase $\omega t$ from the ascending node. Then, in the LVLH frame, we have that the star vector with respect to the satellite is
\begin{align}
    \vu*{s} = [\cos(\phi - \omega t)\sin(\theta),\sin(\phi-\omega t)\sin(\theta),\cos(\theta)].
\end{align}

In 1960, Clohessy and Wiltshire developed a set of linear equations of motion describing the motion of a satellite with respect to another in the LVLH frame \citep{ClohessyW.H.andWiltshire1960}. These equations are an approximation, with no treatment of perturbations and assumptions of circular orbits:
\begin{align}
    \ddot{\rho} &= 3\omega^2\rho + 2\omega\dot{\xi}\\
    \ddot{\xi} &= -2\omega\dot{\rho} \\
    \ddot{\eta} &= -2\omega^2\eta.
    \label{eqHCW}
\end{align}

These equations can be solved yielding $\Delta\vb*{r}(t) = [\rho(t),\xi(t),\eta(t)]$, where
\begin{align}
    \rho(t) &= \kappa_1 + \kappa_2\cos(\omega t-\kappa_3) \label{eq_rho}\\
    \xi(t) &= \kappa_4 - \frac{3\omega}{2}\kappa_1t - 2\kappa_2\sin(\omega t-\kappa_3) \label{eq_xi}\\
    \eta(t) &= \kappa_5\cos(\omega t) + \kappa_6\sin(\omega t) \label{eq_eta}
\end{align}
and $\kappa_1$ through $\kappa_6$ are constants found through the initial conditions.

Now, we see immediately in Equation~\ref{eq_xi} that if $\kappa_1 \neq 0$, then the orbital angular momentum will change and the spacecraft will fly apart. Thus we set initial conditions such that $\kappa_1 = 0$ to remove this motion. Furthermore, suppose that the initial conditions are chosen such that $\kappa_2$ is zero. Then the product $\Delta\vb*{r}\vdot\vu*{s}$ is
\begin{align}
    \Delta\vb*{r}(t)\vdot\vu*{s} = &\kappa_4(\cos(\omega t)\sin(\phi) - \sin(\omega t)\cos(\phi))\sin(\theta) \nonumber\\
    &+ \kappa_5\cos(\omega t)\cos(\theta) + \kappa_6\sin(\omega t)\cos(\theta).
\end{align}
Finally, if we then set 
\begin{align}
    \kappa_5 &= -\kappa_2\tan(\theta)\sin(\phi) & \kappa_6 &= \kappa_2\tan(\theta)\cos(\phi),
\end{align}
we have that $\Delta\vb*{r}(t)\vdot\vu*{s} = 0$ for all $t$. The equations of motion under these constraints are:
\begin{align}
    \rho(t) &= 0 \label{eq_rho2}\\
    \xi(t) &= \kappa_4 \label{eq_xi2}\\
    \eta(t) &= -\kappa_4\tan(\theta)\sin(\phi - \omega t) \label{eq_eta2}
\end{align}

Hence a simple linear array will work for this mission, where the deputy satellites move up and down in the $\vu*{\eta}$ direction with a period equal to that of the orbital period, and with an offset in the $\vu*{\xi}$ direction. In terms of orbits, this means that a configuration where the deputies are in orbits identical to the chief, albeit with a small longitude and inclination offset, will suffice in keeping the array perpendicular to the star at all times. 

\subsection{Key Perturbations}
\subsubsection{J2 Gravitational Perturbation}
While it would be ideal to analytically investigate the theoretical orbits discussed
previously, the spacecraft array will experience perturbations to its orbit due to a
variety of effects. At an altitude of 500 km, the $J_2$ gravitational oblateness is the dominating perturbation. This is a perturbation caused by the oblateness of the Earth, contributing a force in a geocentric frame ($\vu*{x}$ towards the vernal equinox, $\vu*{z}$ towards the Earth's North pole and $\vu*{y}$ completing the orthogonal basis) of 
\begin{multline}
    \vb*{F}_{J_2} = J_2\frac{3\mu R_E^2}{2r^5}\biggl[\left(1-5\frac{z^2}{r^2}\right)(x\vu*{x} + y\vu*{y}) \\+ \left(3-5\frac{z^2}{r^2}\right)z\vu*{z}\biggr],
    \label{3eq_J2_ECI_Force}
\end{multline}
where $R_E$ is the radius of the Earth, $r$ is the radial distance from the centre of the Earth and $J_2$ is the zonal coefficient, approximately 0.00108. This gives an equation of motion in the same frame of:
\begin{multline}
    \ddot{\vb*{r}}_{J_2} = -\frac{\mu}{r^2}\vu*{r} + J_2\frac{3\mu R_E^2}{2r^5}\biggl[\left(1-5\frac{z^2}{r^2}\right)(x\vu*{x} + y\vu*{y}) \\ + \left(3-5\frac{z^2}{r^2}\right)z\vu*{z}\biggr].
    \label{3eq_J2_ECI_acc}
\end{multline}

\begin{figure}
    \centering
    \includegraphics[width=\linewidth]{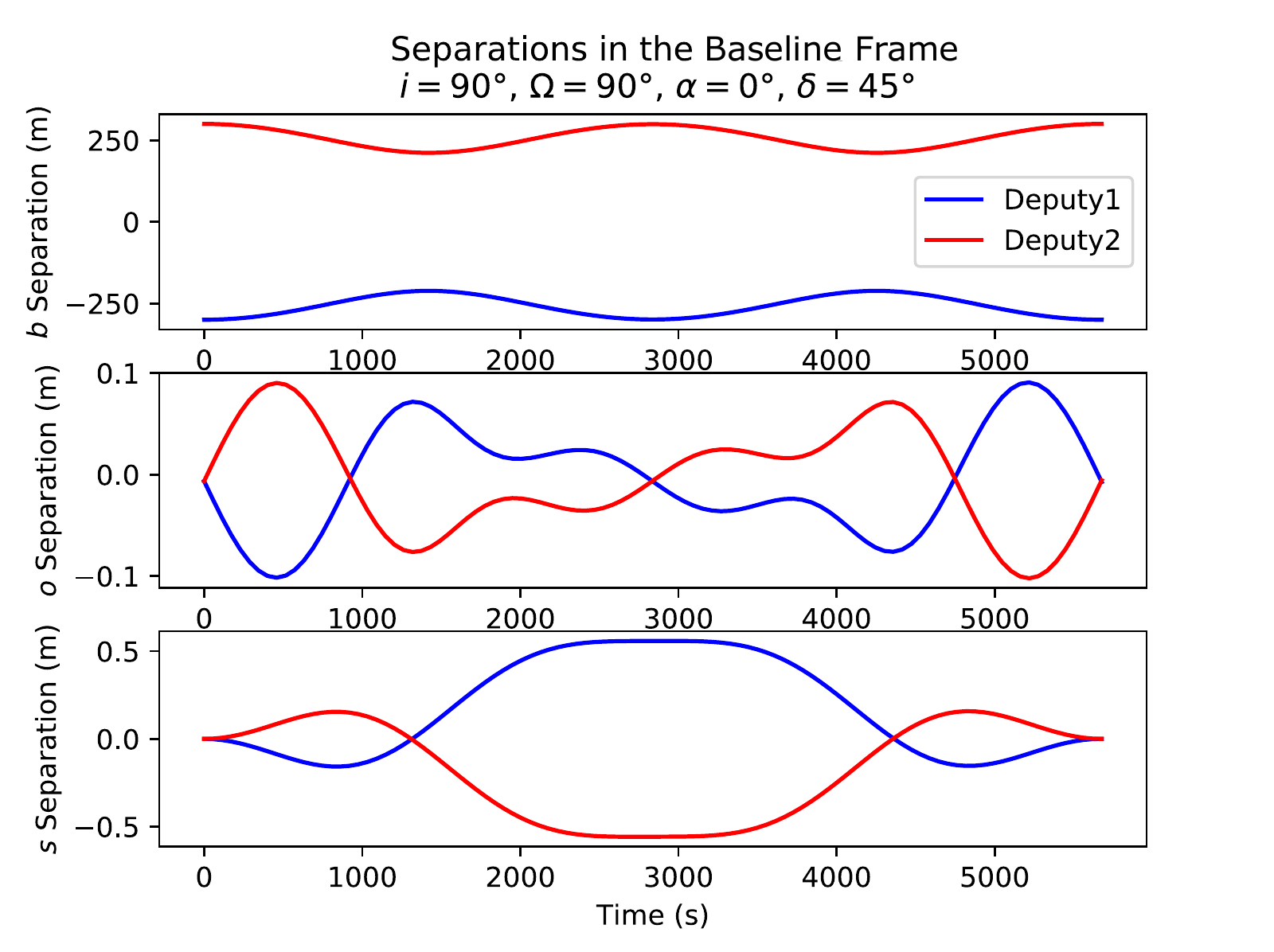}
  \caption{ECI integrated $J_2$ perturbed motion of the two deputy spacecraft with respect to the chief spacecraft over 1 orbit, with a chief orbital configuration of $i=90\degree,\ \Omega = 90\degree$, and star coordinates $\alpha=0\degree,\ \delta = 45\degree$}
  \label{J2}
\end{figure}

Integrating this equation of motion using a 5(4) Runge-Kutta method with a relative error tolerance of 10$^{-12}$ gives us the motions of the satellites under this perturbation.  The motion is plotted in Figure~\ref{J2} for an chief orbital configuration with inclination $i = 90\degree$ and longitude $\Omega = 90\degree$, looking at a star with coordinates $\alpha = 0\degree, \delta = 45\degree$. This motion is plotted in a so called ``baseline frame'' with the $\vu*{b}$ vector being in the direction of the baseline (tangent to the Earth's surface and perpendicular to the star), $\vu*{s}$ being in the direction of the star and $\vu*{o}$ completing the orthogonal basis. 

We see that the $J_2$ perturbation rotates the baseline about $\vu*{s}$ on the order of a fraction of a metre, shown by the motion in the $\vu*{o}$ direction. This twisting is not important for keeping the optical path length identical between the satellites during integration (recall Equation~\ref{eqnOPD}), and so can be safely ignored. The baseline also appears to behave well: the two deputies ``breathe'' in and out in the direction of the baseline, but maintain the same distance from the chief satellite. However, the star ($\vu*{s}$) direction is the most critical to analyse, as it is in this direction that the primary unwanted motion is caused by $J_2$ and we want this separation to be zero. The perturbation causes an unwanted displacement of 0.5\,m; this small amount should be able to be removed through thrusting during an integration and will be explored later.

\subsubsection{Atmospheric Drag}
\begin{figure}
  \centering
  \includegraphics[width=0.7\linewidth]{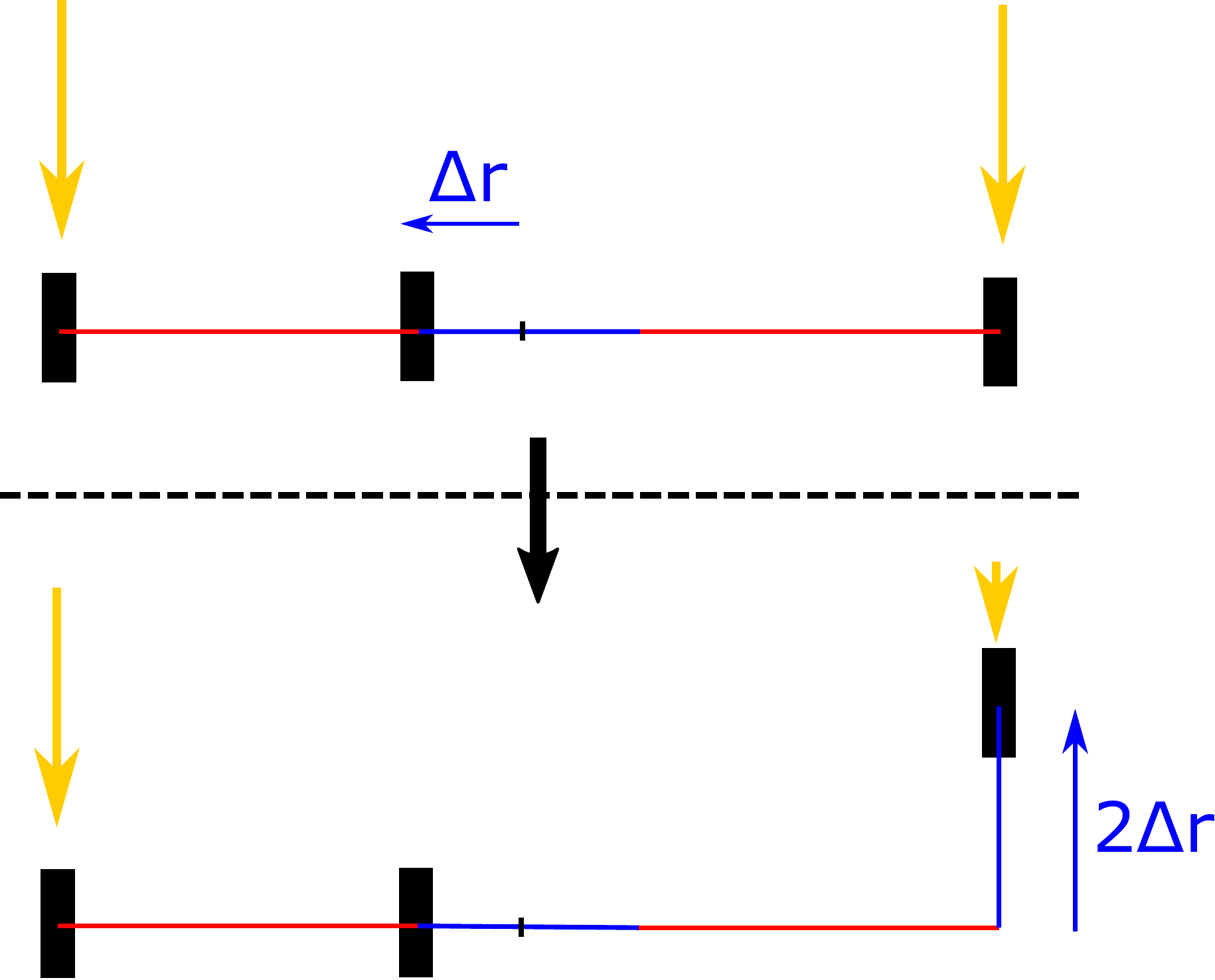}
  \caption{Potential routine for correcting for drag. In the top panel, drag has caused the chief to move a distance $\Delta r$ from the centre of the array. To correct for this, we could boost the deputy with the longer separation up $2\Delta r$ in the direction of the star. Essentially, compensating an increase in $|\Delta\vb*{r}|$ with an increase in $\Delta\vb*{r}\vdot\vu*{s}$, so that $D$ is kept close to 0 (see Equation~\ref{eqnOPD}).}
  \label{drag}
\end{figure}

Atmospheric drag will also contribute to perturbing the orbits, albeit less importantly than $J_2$. We calculate that drag should, at worst, differentially affect the spacecraft array by moving the chief satellite closer to one of the deputies than the other with an acceleration of about $1\times 10^{-7}$\,ms$^{-2}$ (as in Figure~\ref{drag}). Over half an orbit ($t = 2700$\,s), we then get a displacement of
\begin{align*}
    \Delta r &= \frac{1}{2}at^2\\
    &\approx 35 \text{\,cm}.
\end{align*}

As the drag displacement shortens one path and lengthens the other (see Figure~\ref{drag}), we must compensate for twice the displacement to make the paths equal. Thus, to correct a change in separation of this magnitude, we could thrust one of the deputies up by 70\,cm.  Over a 300\,m chief/deputy separation, this would correspond to an angular change of 0.1\degree, and so would remain in the field of view of the chief satellite. Adjusting for this small angular change should also be achievable with an actuator attached to the mirror. 

Atmospheric drag will also allow the satellite system to naturally deorbit after the mission has ended. We calculate that the characteristic time taken to descend one scale height of the atmosphere at 500\,km is about 4.75 years. This shows that we should be able to both comfortably achieve our desired mission length of three years, while also deorbiting long before the guideline deorbit time of 25 years \citep{EuropeanSpaceAgency2015}. 

\subsection{Correcting the Perturbations}
\begin{figure*}[t!]
    \centering
    \includegraphics[width=0.6\linewidth]{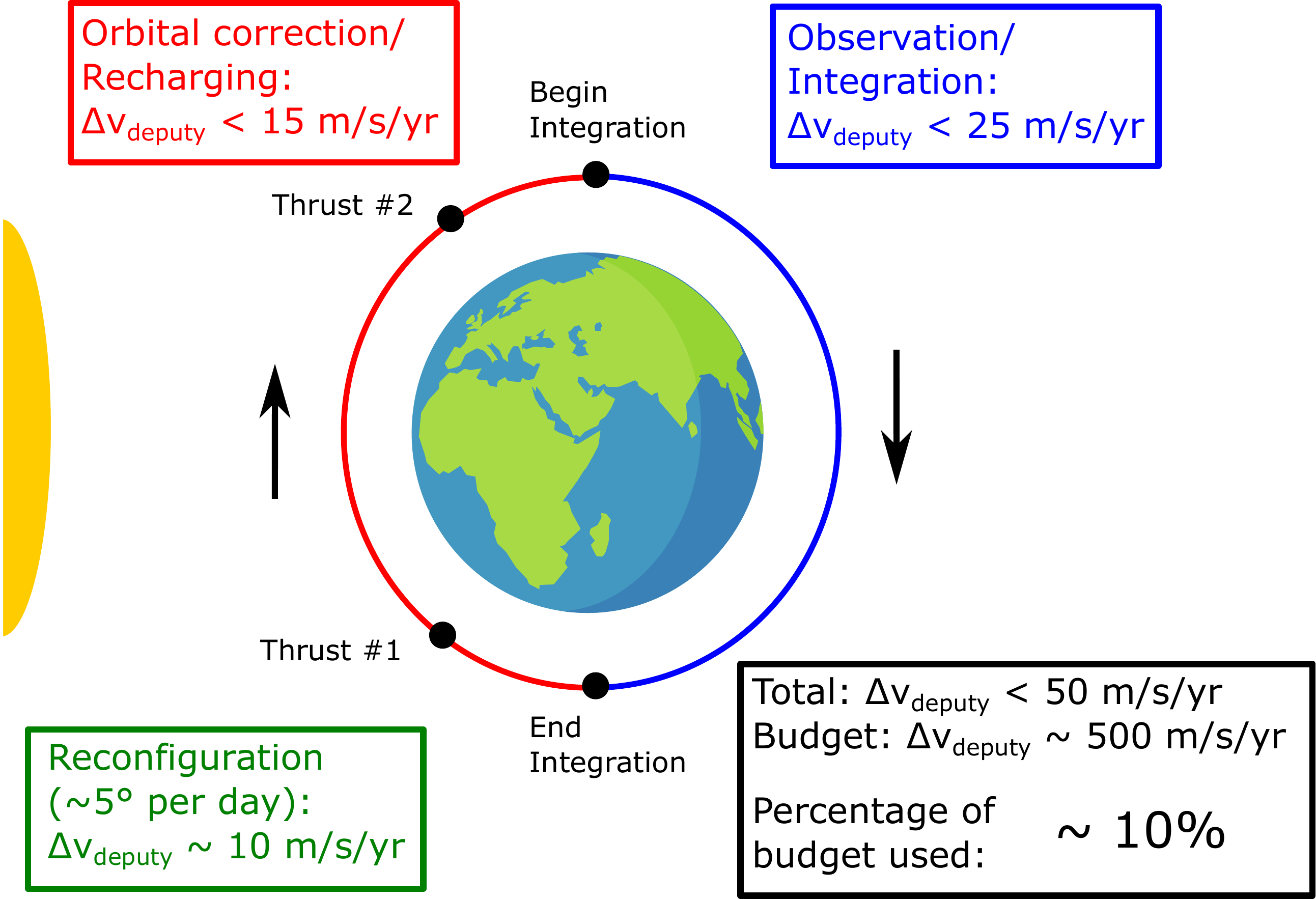}
    \caption{Summary of the $\Delta v$ required by a deputy satellite to correct for orbital perturbations over a year.}
    \label{corrections}
\end{figure*}

Previously, we identified how the satellite array will be perturbed from its original orbit due to the influence of the $J_2$ geopotential term. What is more significant, however, is how to correct these perturbations. There are three categories of orbital corrections we need to consider. The first two detail the corrections needed during a regular orbit: one during integration (i.e. when the star is being observed) and the other correcting the orbit during the solar recharge phase (when the array is on the daylight side of the Earth). The final type of orbital manoeuvre are those that are required for when the target of interest or maximum baseline is changed.

In each correction, the most important parameter is $\Delta v$, which is the change in velocity due to an impulsive manoeuvre \citep{Curtis2014}. The $\Delta v$ budget is used as a stand in for the amount of fuel required.
As an example budget, we consider the \cite{EnpulsionSpacecraftTechnology2018IFMThruster} microthruster, which has a total budget of 5000\,Ns. For a 3U CubeSat at 3\,kg (such as the deputies), a single thruster would provide approximately $\Delta v \approx 2$\,km/s in total over the lifetime of the thruster, and for a 6\,kg CubeSat (such as the chief), this would decrease to about 1\,km/s.  Assuming a mission length of three years and a period of 90\,min (orbital altitude of 500\,km), we find that, per orbit, we have a $\Delta v$ budget of $\sim$0.1\,m/s for the deputies and $\sim$0.05\,m/s for the chief. Multiple simulations were conducted to identify the $\Delta v$ requirements, and is summarised for a deputy satellite in Figure~\ref{corrections}.

Firstly, we will consider correcting the perturbations during the period of integration - when the array is looking at the star. This is for approximately the first half of the orbit, while the array is on the nighttime side of the Earth (the blue portion of Figure~\ref{corrections}). During this time the optical path difference needs to be kept close to zero (Equation~\ref{eqnOPD}). This is the main requirement, and so we will only focus on correcting the components of the optical path difference rather than the total position and velocity. 

The path difference can be split into two components: the baseline separation and the star separation. The baseline separation is given by:
\begin{align}
    \Delta b = |\Delta\vb*{r}_1| - |\Delta\vb*{r}_2|.
\end{align}
We also calculate the component of the separation of deputy $j$ in the star's direction, $\Delta s_j$:
\begin{align}
    \Delta s_j &= \Delta\vb*{r}_j\cdot\vu*{s}
\end{align}

If we take the time series of these values over the course of an orbit, we can then calculate the $\Delta v$ required to counteract the perturbations. This is calculated for a separation $\Delta x$ below:
\begin{align}
    \Delta v_x &= \int |\ddot{\Delta x}|
\end{align}
where $\ddot{\Delta x}$ can be found by taking the numerical gradient of the time series twice.

The above way that $\Delta v$ is calculated assumes that the thrust direction can be in any direction, but under the baseline thruster configuration, all three satellites can only thrust upwards in the $\vu*{s}$ direction. Despite this, we can still make an approximation that should give us an upper bound on the required $\Delta v$. Considering the $\vu*{s}$ direction, if the separation is negative then the satellite must thrust in a positive direction. If the satellite has a positive separation with respect to the chief, then rather than thrusting negatively, we can force the other two satellites to thrust positively. Hence, the deputies will thrust at most $2\Delta v_{s}$: once for their own positive thrusts and once for the other deputy's negative thrusts. Conversely, the chief will only require one of these thrusts - the perturbations on the two deputies should be symmetrical about the chief, and so the combined ``negative'' thrusts of the deputies should equal $\Delta v_{s}$.

For the baseline direction, we can compensate a change in the baseline direction with a shift in the $\vu*{s}$ direction. As we cannot disentangle which of the deputies has the larger separation at a given point, we will take a worst case scenario and add the $\Delta v$ in this direction to both of the deputies. Therefore we have that the maximum $\Delta v$ for all three satellites is
\begin{align}
    \Delta v_1 &< 2\Delta v_{s} + \Delta v_b\\
    \Delta v_2 &< 2\Delta v_{s} + \Delta v_b\\
    \Delta v_c &< \Delta v_{s}. 
\end{align}

Using the numerical method described above, we find that we require a $\Delta v$ of approximately 25\,m/s/yr to counteract $J_2$ and atmospheric drag.

We also need to identify the $\Delta v$ required to move the array back into position during the daytime half of the orbit. Though we do not have the optical path constraint as before due to the fact we are not taking measurements, we have a secondary constraint in that we must have ample time to charge the array's batteries. Hence, we need to limit the time that the satellites are active and allow the array to passively charge for a majority of the orbit.

We therefore propose to limit the number of thrusts to two, but optimise them so that they take the spacecraft as close to the beginning of the orbit as possible (the red part of Figure~\ref{corrections}). In each orbit, we model conducting one thrust at 10 minutes after finishing the integration, followed by a coasting phase (while batteries are recharged), then we conduct a second thrust 10 minutes before restarting integration. These thrusts can be in any direction as we can change where each spacecraft is pointing during this portion of the orbit.

Let us consider that after two thrusts, deputy $j$ has final state vector $\vb*{X_{f}}_j = [r_{bj},r_{oj},r_{sj},\dot{r}_{bj},\dot{r}_{oj},\dot{r}_{sj}]$ in the baseline frame. Let us then say that the desired end state vector is $\vb*{Y_{f}}_j$. The residuals are then:
\begin{align}
    \vb*{\Delta X_{f}}_j = \vb*{X_{f}}_j - \vb*{Y_{f}}_j. \label{4eq_root}
\end{align}
Since we have six unknowns per deputy satellite, and the end state has six parameters, we can explicitly solve for the thrusts such that the residuals are zero: $\vb*{\Delta X_{f}}_j = 0$. Numerically solving this results in a $\Delta v$ requirement of 15\,m/s/yr.

Finally, to factor in reconfiguring the orbit, we also assume that a correction of $\sim 5\degree$ per day is sufficient. {\red Section~\ref{secAstrophysics} describes example observing strategies that use of order this slew rate.} 
This amount of reconfiguration results in a $\Delta v$ requirement of approximately 10\,m/s/yr.

Combining the three types of corrections, we find that the deputy satellites will use roughly 50\,m/s per year, which is 10\% of the yearly $\Delta v$ budget that a single thruster would provide (500\,m/s/yr). This is a very useful result, showing that the one thruster model should be sufficient and that we have a large amount of headroom for other thrust requirements, such as the control system and emergency manoeuvres.

\subsection{Observability}
\begin{figure}
  \centering
  \includegraphics[width=0.75\linewidth]{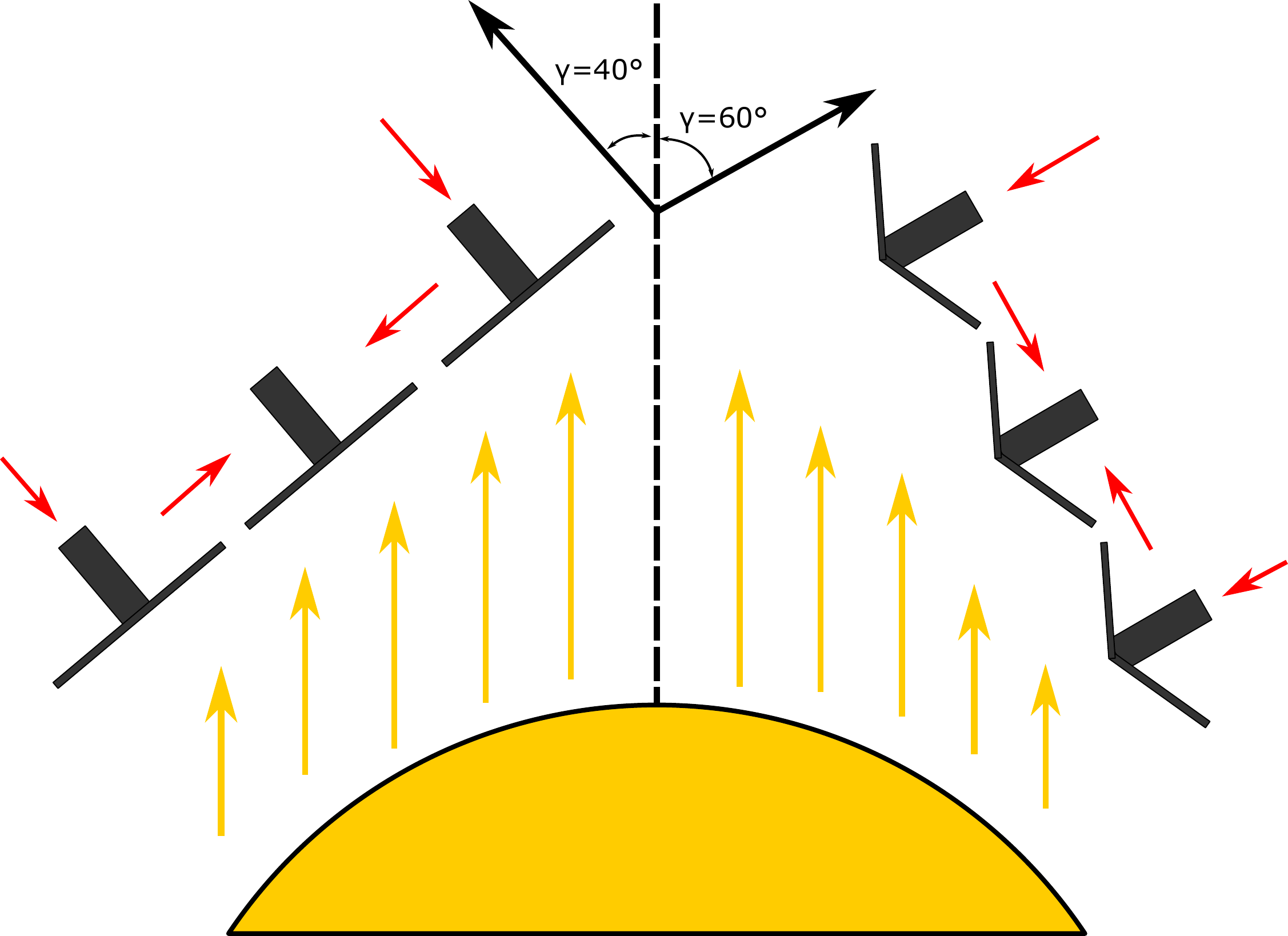}
  \caption{Schematic of keeping the satellite array within 40/60\degree\ of the antisolar axis. Note that the solar panels can be used to baffle the sun, even for high antisolar angles $\gamma$, by folding them inwards slightly.}
  \label{antisolar_angle}
\end{figure}

Another key part of this feasibility study was to identify the amount of sky coverage that an array in LEO could achieve. We identified two main factors that would prevent a measurement from taking place:

The first is whether the Sun-object angle prevents observation, as characterised by the antisolar angle $\gamma$. Both the telescope structure and the beamline between the chief and deputy must be in shadow.  A schematic of this is in Figure~\ref{antisolar_angle}. We want to have a wide angle in order to see as much of the sky as possible. However, for our feasibility study design, we note that $\gamma = $ 60\degree\ antisolar is the upper limit for how wide we can go with easily deployable baffling. 

The second major factor is whether the Earth is obscuring the star. When factoring this into the simulation, we extended the radius of the Earth slightly to account for the glow of the Earth's atmosphere, which would block a measurement. There is a potential third criteria, in whether light from the Earth will interfere with the light path between the spacecraft. However, for our orbital configuration this will not be an issue as the array should always be tangential to the surface of the Earth. Here, we note that the reflected light from the Moon is generally not important as long as we restrict science targets to a magnitude of $\sim m_V < 15$. Finally, we don't consider occultations by the moon, as they are generally short and only block a small area of the sky ($\sim 0.5\degree$).

\begin{figure}[t!]
\begin{subfigure}{\columnwidth}
    \centering
    \includegraphics[width=\linewidth]{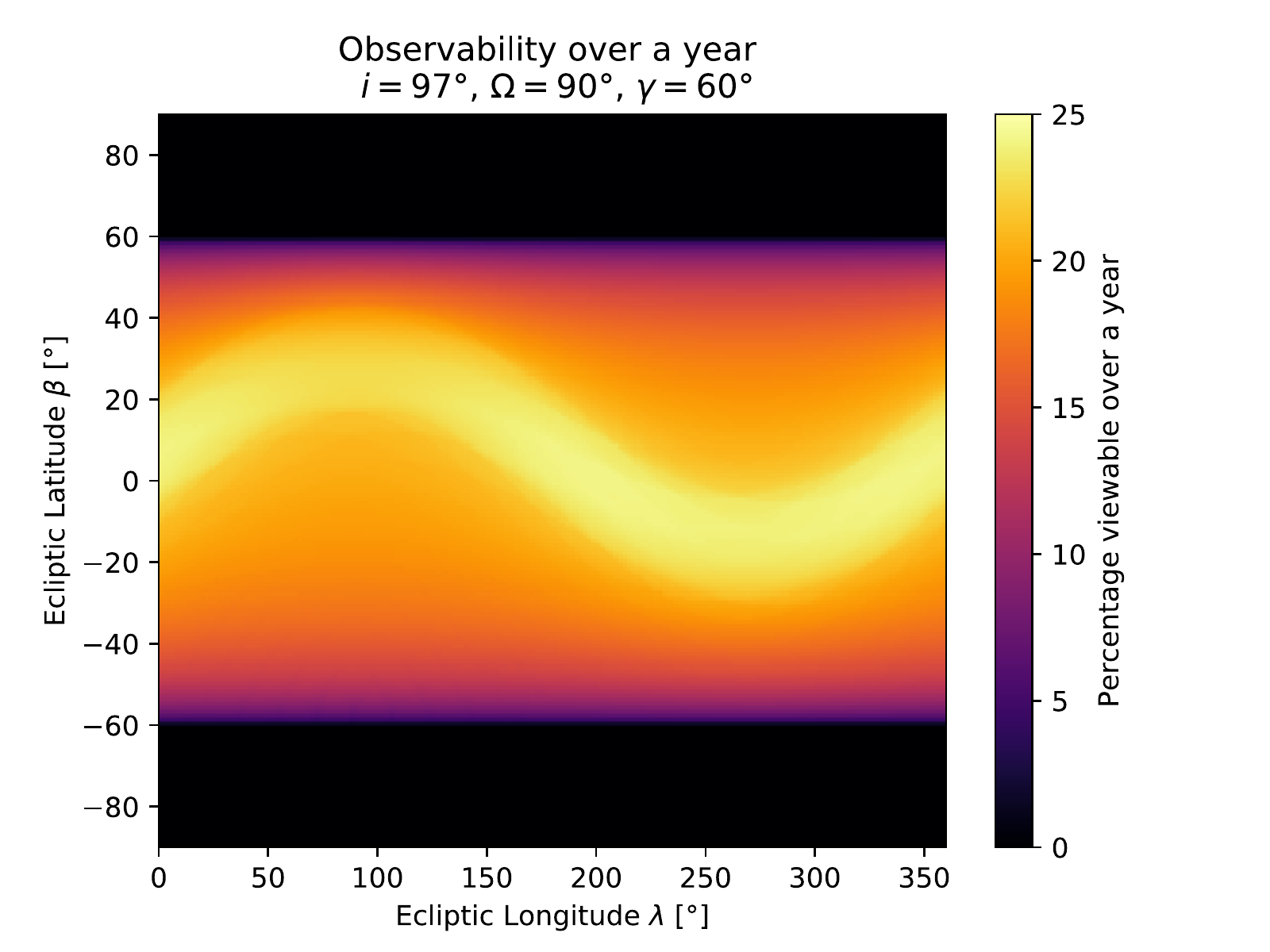}
  \caption{Longitude $\Omega = 90\degree$, Inclination $i = 97\degree$\\ (heliosynchronous orbit)}
  \label{obs_map_helio}
\end{subfigure}
\begin{subfigure}{\columnwidth}
    \centering
    \includegraphics[width=\linewidth]{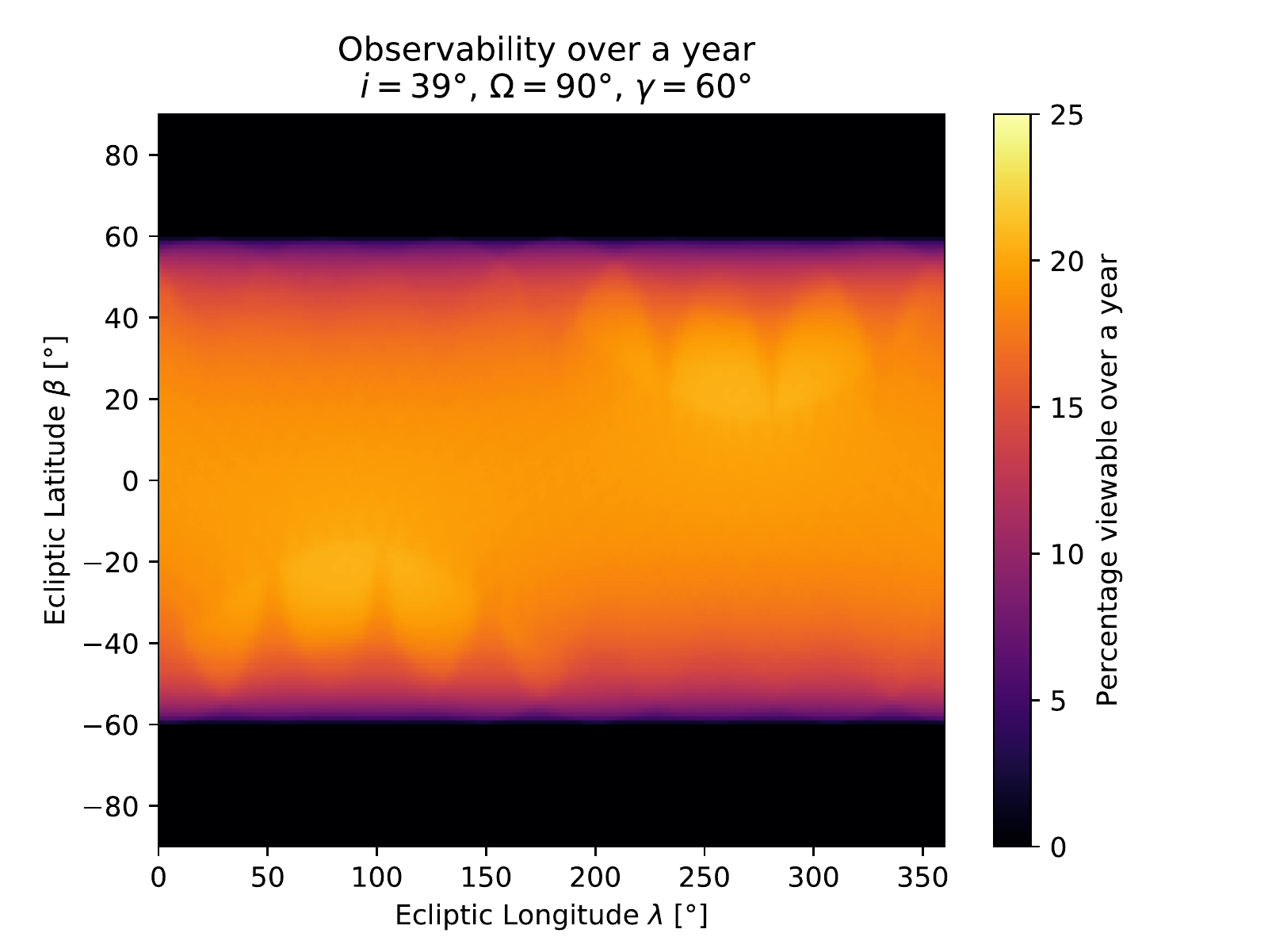}
  \caption{Longitude $\Omega = 90\degree$, Inclination $i = 39\degree$}
  \label{obs_map_nz}
\end{subfigure}
\caption{Map of the satellite sky coverage for different orbital configurations, as a percentage observable over a year, shown in ecliptic coordinates. The antisolar angle was set at 60\degree\ for both plots.}
\label{obs_map}
\end{figure}

Together, a simulation was run to determine the proportion of the sky that was viewable over a year's worth of orbits. A map in ecliptic coordinates, describing the percentage viewable of a coordinate over a year is found in Figure~\ref{obs_map} {\red for two example orbital configurations. Figure \ref{obs_map_helio} shows a dawn/dusk heliosynchronous orbit, with an inclination of $i=97\degree$ and longitude of $\Omega = 90\degree$ from the vernal equinox, while Figure \ref{obs_map_nz} has an inclination of $39 \degree$. The heliosynchronous orbit was chosen as it is both a common orbit for Low Earth Orbit satellites, and allows targets that are at opposition from the sun to be observed throughout the whole orbit, rather than half an orbit as described in Figure \ref{corrections}. This is reason for the bright stripe in the plot. However, despite this advantage, half an orbit will still be required to correct for orbital perturbations and so will have an integration time of approximately 45~minutes. The 39\degree\ orbit was chosen as it is the lowest inclination that the rocket company Rocket Lab can achieve \citep{RocketLab2019PayloadGuide}, therefore representing the inclination of the highest launch mass. Furthermore, as the satellite array will be assembled in Australia, the proximity to New Zealand results in this being the least expensive custom launch option.}

We can see in the plots that the antisolar angle sets the ecliptic latitude at which we can view targets - stars with a latitude greater than the antisolar angle will not be viewable. Furthermore, we see that for targets with latitudes $-40\degree\ < \beta < 40\degree$, they are viewable for between {\red 15} - 25\% of the year. 

\subsection{Enabled Astrophysics}
\label{secAstrophysics}

Although this paper focuses on the technical requirements of a prototype, it is worth considering the astrophysics enabled by such a mission. At a CubeSat scale, the aperture diameter is no larger than Fried's coherence length in the visible, which means that increased sensitivity only comes from the increased coherence time. This is, however, quite substantial: of the order of 1000 in-between the typical 2--5\,ms coherence time from the ground and an anticipated 2--5\,s exposure time in space. The simplest type of instrument, as briefly described in Section~\ref{refArrayDesign}, is an unpolarised moderate resolution spectrograph recording fringes. Given that the science beam combiner must also act as fringe tracker, visibility and differential phase are the key observables. Differential phase is especially relevant to imaging of H-$\alpha$ where there is a continuum source which is independently known to be compact.

Some appealing targets that lie within 40\degree\ of the ecliptic are the $\rho$ Ophiuchus and Taurus star forming regions that could be used to analyse stars in the process of formation. Other objects include the microquasar SS 433, a black hole of a few solar masses located inside a supernova remnant and exhibiting bright emission in the radio, optical and X-ray \citep{Margon1984Observations433}. 3C 273 is also a potential target, which was the first quasar ever discovered \citep{Schmidt19633CRed-Shift} and is the optically brightest in the sky. All of these have a magnitude of V $< 15$.

In addition to this astrophysics enabled by differential quantities, such an interferometer could of course be used for traditional measurements based on the visibility modulus. This could include precision orbits of binaries detected with Gaia, and precision diameters including the interferometric Baade-Wesselink method. {\red Precision calibration of surface brightness remains an important astrophysical goal, with renewed relevance for cosmology \citep{Mould19,Riess19}.} We argue that  closer examination of these myriad science cases is needed to determine if the class of space interferometer presented here is best thought of as a prototype or a science mission in its own right. 

\begin{figure}
    \centering
    \includegraphics[width=\columnwidth]{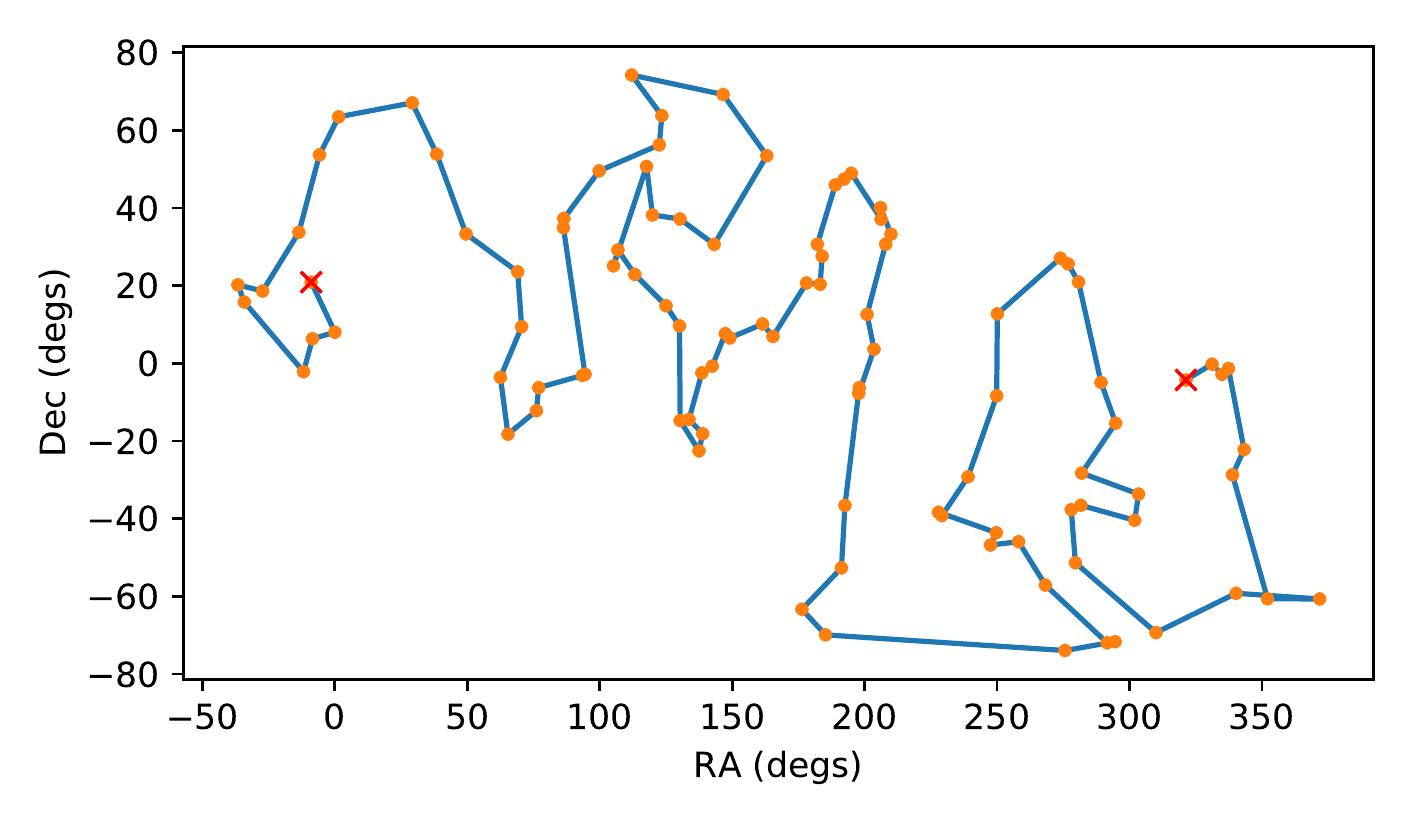}
    \caption{An example optimised path through 100 targets observed over 1 year, minimising total slew time to 3.4 degrees per day on average while maintaining the anti-solar angle constraint.}
    \label{figAllSky}
\end{figure}

{\red Given that reconfiguration slew rate (degrees per day) is a key mission parameter, we can consider different observing scenarios and their impact on the science mission. As an example, Figure~\ref{figAllSky} shows an observing sequence of 100 objects over 
a year, selected from a target list of 150 objects uniformly distributed in the sky. The
target list included objects in the unobservable ecliptic poles, and used a limiting anti-sun angle $\gamma$
of 60 degrees. The chosen path for this 
travelling salesman problem was found via simulated annealing. A mean slew distance of 3.4 
degrees per day is needed in this example, corresponding to an average of 12.5 degrees 
between targets. Increasing the number of targets 
to 360 increases the slew per day to 8.6 degrees, while restricting the targets to within
15 degrees of the Galactic plane, applicable to Galactic targets at distances greater than $\sim$ 100\,pc, reduces the mean slew distance to 6 degrees between targets or 1.7 degrees per day.}

%% file: 4Conclusion.tex
\section{Conclusion and Future work}

In order to analyse exoplanet formation and atmospheres, we require instruments with extremely high resolution and sensitivity; a level at which current instruments fall short. Seemingly, the only future types of instruments that will satisfy both of these requirements is an interferometer located in space, which brings a whole host of technical challenges. Missions from large organisations such as TPF-I from NASA and \textit{Darwin} from ESA were cancelled, leaving a technological gap that has remained since the mid 2000s. 

Here, we have developed and tested the feasibility of a small scale space-interferometry mission that aims to restart developments in the field of space interferometry. The design of this mission is ambitious: two 3U CubeSats that act as telescopes, and one 6U beam combiner CubeSat located between them. With such a space constraint, and to reduce cost, a single thruster for each satellite would be ideal. After running some orbital simulations, we find that the $\Delta v$ required to counteract the $J_2$ and atmospheric drag perturbations only required $\sim$10\% of the available $\Delta v$ budget per year; an encouraging result that shows that three thrusters should be enough to perform interferometry with a large amount of headroom.

We also find with visibility simulations that the antisolar angle is a critical parameter in determining the percentage of the year that a target is viewable. Assuming an angle of 60\degree, we find that targets with ecliptic latitudes $-40\degree\ < \beta < 40\degree$ are viewable for between 20 - 25\% of the year. 

As a prototype mission, the next stage is to define exactly what is to be prototyped for future major exoplanet missions, and ensure that appropriate technological readiness levels can be reached. This is likely to include interspacecraft metrology, formation flying control to an adequate level in both angle and position with more than 2 spacecraft, as well as {\red adequate fringe tracking using a combination of science fringes and metrology}: to our knowledge none of these have achieved {\red full technology readiness for space} at this point. 

Finally, as noted in section \ref{secAstrophysics}, there are a range of astrophysical goals for a low earth orbit mission that are competitive with ground-based technologies. If astrophysics is to be a goal of this kind of mission, core science requirements also need to be identified from an appropriately broad collaboration.